\documentclass[11pt,a4paper]{article}
\usepackage[utf8x]{inputenc}
\usepackage[T1]{fontenc}
\usepackage{anyfontsize}
\usepackage[english]{babel}
\usepackage{amsfonts}
\usepackage{eucal}
\usepackage{amsmath}
\usepackage{color}
\usepackage[colorlinks=true, linkcolor=black,citecolor=blue, urlcolor=blue]{hyperref}
\usepackage{tikz}
\usetikzlibrary{shapes,arrows}
\tikzstyle{block} = [rectangle, draw, fill=blue!20, 
    text width=6em, text centered, rounded corners, minimum height=2em]
\tikzstyle{io} = [rectangle, draw, fill=red!20, 
    text width=4.5em, text centered,node distance=3cm, minimum height=2em]
\tikzstyle{line} = [draw, -latex']
\usepackage{mathptmx}

\def\one{\mbox{1\hspace{-4.25pt}\fontsize{12}{14.4}\selectfont\textrm{1}}}

\newcommand\Tstrut{\rule{0pt}{2.6ex}}         
\newcommand\Bstrut{\rule[-0.9ex]{0pt}{0pt}}

\title{Automatic cry analysis and classification for infant pain assessment}

\author{Davide Ricossa \thanks{e-mail: davide.ricossa@edu.unito.it; Dipartimento di Matematica "G. Peano", Universit\`{a} degli Studi di Torino, Via C. Alberto 10, Torino, Italy;} 
              \and  
              Enrico Baccaglini \thanks{e-mail: baccaglini@ismb.it; MLW, Istituto Superiore Mario Boella, Via P. C. Boggio 61, Torino, Italy;}
	          \and 
	          Elvira Di Nardo \thanks{e-mail: elvira.dinardo@unito.it; Dipartimento di Matematica "G. Peano", Universit\`{a} degli Studi di Torino, Via C. Alberto 10, Torino, Italy;}
	          \and 
	          Emilia Parodi\thanks{e-mail: emilia.parodi@unito.it; SC di Pediatria e Neonatologia, AO Ordine Mauriziano, Largo F. Turati 62, Torino, Italy;} 
	          \and 
	          Riccardo Scopigno \thanks{e-mail: scopigno@ismb.it; MLW, Istituto Superiore Mario Boella, Via P. C. Boggio 61, Torino, Italy;}
	          }

\date{}

\begin{document}

\maketitle

\begin{abstract}
The effectiveness of pain management relies on the choice and the correct use of suitable pain assessment tools. In the case of newborns, some of the most common tools are human-based and observational, thus affected by subjectivity and methodological problems. Therefore, in the last years there has been an increasing interest in developing an automatic machine-based pain assessment tool.

This research is a preliminary investigation towards the inclusion of a scoring system for the vocal expression of the infant into an automatic tool. To this aim we present a method to compute three correlated indicators which measure three distress-related features of the cry: duration, dysphonantion and fundamental frequency of the first cry. In particular, we propose a new method to measure the dysphonantion of the cry via spectral entropy analysis, resulting in an indicator that identifies three well separated levels of distress in the vocal expression. These levels provide a classification that is highly correlated with the human-based assessment of the cry.
\end{abstract}
\newpage

\section{Introduction}
\begin{table}[t]
\centering
\caption{Douleur Aigu\"{e} du Nouveau-né}
\scalebox{0.8}{
\begin{tabular}{l | c}
\hline
 Indicators & Score \Tstrut \\
\hline
\textbf{Facial expression: }\textit{eye squeeze, brow bulge, nasolabial fold;} &  \Tstrut \\
\hspace{0.5cm} Calm & 0 \\
\hspace{0.5cm} Snivels and alternates gentle eye opening and closing & 1 \\
\hspace{0.5cm} Mild, intermittent with return to calm & 2 \\
\hspace{0.5cm} Moderate & 3 \\
\hspace{0.5cm} Very pronounced, continuous & 4\Bstrut\\
\hline
\textbf{Limb movements:} \textit{pedals, toe spread, legs tensed and} & \Tstrut \\
\hspace{0.5cm}\textit{pulled up, agitation of arms, withdrawal reaction;} & \\
\hspace{0.5cm} Calm or gentle & 0 \\
\hspace{0.5cm} Mild, intermittent with return to calm & 1 \\
\hspace{0.5cm} Moderate & 2\\
\hspace{0.5cm} Very pronounced, continuous & 3\Bstrut \\
\hline
\textbf{Vocal expression} & \Tstrut \\
\hspace{0.5cm} No complaints & 0\\
\hspace{0.5cm} Moans briefly & 1 \\
\hspace{0.5cm} Intermittent crying & 2\\
\hspace{0.5cm} Long-lasting crying, continuous howl & 3\Bstrut\\
\hline
\end{tabular}
}
\end{table}
Until the '80s, due to the lack of scientific studies, there was just a set of assumptions about infant pain, which resulted in a common undertreatment of it. Among these assumptions, the major one was that infants do not experience pain due to their neurological immaturity. This assumption was later proven to be incorrect \cite{anand1987pain}. Moreover, the number of painful events increases with the most immature infants. This fact, jointly with the awareness of the short- and long-time adverse sequelae of the exposure to repeated painful stimuli in early life \cite{hermann2006long}, has made the infant pain assessment a real issue.  
Nowadays there are numerous neonatal pain scales, that use some observable indicators as surrogate of the patient's self-evaluation. An example is the  Douleur Aigu\"{e} du Nouveau-né (DAN) scale \cite{carbajal1997dan}, reported in Table 1. Thus, due to the observational nature of these scales, it is difficult to identify a peak of pain in an acute pain experience and a continuous assessment is not applicable for chronic pain. Besides these methodological problems, human-based tools can also be affected by subjectivity problems. For instance, Bellieni et al. \cite{bellieni2007inter} observed a significant difference between three groups of operators (O1, O2 and O3) using  some of the most common tools to assess the pain of infants undergoing a routine heel prick procedure as follows: O1 scored after performing the actual heel prick, O2 scored as an observer who was free to watch the procedure closely, O3 recorded the procedure through a video camera and gave the score later by watching the video more than once if necessary.
Because pain is subjective \cite{merskey1994classification}, a second degree of subjectivity is added and so the  bias of a human observer could really compromise the reliability of the pain assessment process. This is why in the past several years there has been an increasing interest for an entirely machine-based pain assessment tool \cite{zamzmi2016machine}: a way to monitor automatically the various pain indicators and evaluate them continuously and consistently with a minimum bias. 

\section{Background}
\subsection{Infant cry analysis}
Crying is the earliest form of communication which constitutes the major part of the infant's vocalization: it is the way the newborn expresses his/her physical and emotional state and needs. Today's research in infant cry
analysis was initiated by a team of Scandinavian researchers in the '60s who proposed spectrographic analysis \cite{wasz1985twenty} as one of the first approaches. Later, with the development of high-speed computer technology the study of cry has been subject to significant improvement.
Over the years, the investigation of the main characteristics of infant cry both in time and in frequency domain has brought to light important insights related to the cry generation process and some models have been proposed \cite{golub1985physioacoustic,facchini2005relating}. Nevertheless, our knowledge of cry generation is still limited. Therefore, nowadays infant cry is mainly studied from the processing \cite{orlandi2015application}. Being the product of a human's vocal apparatus, although an immature one, it can be considered a particular case of human voice. Studying infant cry using speech signal processing/recognition techniques can thus be a promising approach. These techniques have led to identify some time-frequency patterns\footnote{Though, some of these patterns have been just described verbally in literature, rather than defined numerically.}, or crying features, that are correlated with the context and, therefore, are considered meaningful. For instance, following a linguistic approach, in 1993 Xie et al. \cite{xie1993determining} first defined a set of 10 cry phonemes which provide the basis of the major part of the time-frequency pattern of variation in infant cry. Then, they analysed the correlation between the permanence time in each cry mode and the level of distress (LOD) perceived by the parents. It was observed \cite{xie1993determining} that amongst the phonemes, the dysphonation shows the most consistent positive correlation with the perceived LOD. An other example is the fundamental frequency, which is considered an outstanding characteristic \cite{baeck2001study}. Indeed, it has been observed that the first cry produced in response to an invasive pain stimulus displays a higher fundamental frequency and greater variability in the fundamental frequency during the cry episode \cite{porter1986neonatal}.

On the other hand, cry interpretation is still a difficult task. In fact, different crying features give information about the LOD of the infant, rather than reflecting the exact reason for crying (e.g. hunger, discomfort, loneliness, pain, colic pain etc.) which is contextual. So, an observed cry helps to identify a set of possible causes or stimuli, but each of them has to be taken as uncertain and presumed without more information about the scene \cite{bellieni2004cry}. Over the last 30 years, several models for crying classification have been proposed, with various results (e.g. Hidden Markov \cite{xie1996automatic} and Gaussian Mixture-Universal Background \cite{buanicua2016automatic} models, Bayesian \cite{baeck2003bayesian} and Random Forest \cite{orlandi2015application} classifiers). Moreover, because of the absence of an actual ground truth for cry interpretation (quite often the context, e.g. "distressful" or "not distressful", is used in place of it), it is still unclear which one is the best performing approach (see Section 6). Some proposed classifications are strictly related to the processing technique used to analyse the cry, which sometimes relies on the use of a non-fully-accessible (non-free or experimental) software, thus making the comparison even more difficult.
\subsection{Cry Analysis for Infant Pain Assessment}
A preliminary step towards the inclusion of the vocal expression into a machine-based pain assessment tool is to understand if the acoustic features of a painful cry fit some kind of scoring system. To this end, in this paper we present a method to evaluate three indicators of some distress-related features of the cry. Then, we analyze the indicators' correlation when calculated for a dataset of infants subjected to procedural pain. Finally, we compare the results with the sample mode of the human-based assessment of the infant's vocal expression.

The rest of the paper is structured as follows. Section 3 describes the experimental setting and the dataset used in the experiment. Section 4 presents the pre-processing and the proposed method to calculate the indicators. Section 5 resports the experimental results, which are later discussed in Section 6.\newpage

\section{Design}
\subsection{Project}
This paper is part of a preliminary investigation commissioned by AO Ordine Mauriziano di Torino (Italy) to Istituto Superiore Mario Boella. Given a small amount of data, the aim of the project was to explore machine-based methods for infant pain-assessment in order to study the  feasibility of an automatic pain assessment tool. This is a mandatory step in order to ask for the ethical committee to approve a clinical trial involving a bigger cohort of subjects.
\subsection{Subjects and data acquisition}
This study is based on the analysis of a cohort of 31 healthy term infants who underwent heel lance for neonatal screening. The heel lance, is a compulsory procedure (L. 104/92 - art. 6 \cite{legge10492}) that must be performed on every newborn between the first 48 and 72 hours after birth before the hospital discharge. During the procedure, the heel of the newborn is first lanced and then is gently squeezed in order to soak a blood sample into pre-printed collection cards. A monitoring system\footnote{An AXIS M1034-W network camera fixed to the wall through a mounting bracket and connected to a PC. Quality HDTV 720p/1 MP. The resolution varies from 1280x800 to 320x240 pixels and the frame rate is 25/30fps. The sampling rate of the audio signal is 16 kHz.} recorded the reaction of each newborn in the course of the procedure. For every infant in the dataset a parent has given formal consent for the audio-video recording and every recommended measure has been taken so as to minimize the intervention-related stress and pain.
\section{Methods}
\begin{figure}[t]
\centering
\begin{tikzpicture}[node distance=1cm]

\node[block] (FS) {First \\ Segmentation};
\node[io, left of= FS] (in1) {$C_{i}, N_{i}$};
\node[io,right of=FS] (o1) {$SC_{i}^{0}, \overline{r}_{i}$};
\path [line] (in1) -- (FS);
\path [line] (FS) -- (o1);

\node[block,minimum height=2em, below of=FS] (B) {Bootstrap};
\node[io, left of= B] (in2) {$\{SC_{i}^{0}\}_{i=1}^{N}$};
\node[io,right of=B] (o2) {$\hat{q}_{.85}$};
\path [line] (in2) -- (B);
\path [line] (B) -- (o2);

\node[block, below of=B] (SFE) {Final Segmentation};
\node[io, left of= SFE] (in3) {$SC_{i}^{0}$, $ \hat{q}_{.85}$, $\overline{r}_{i}$};
\node[io,right of=SFE] (o3) {$SC_{i}$};
\path [line] (in3) -- (SFE);
\path [line] (SFE) -- (o3);

\node[block,  node distance=1.1cm, minimum height=2em, below of=SFE] (E1) {Feature Extraction I};
\node[io, left of= E1] (in4) {$SC_{i}$};
\node[io,right of=E1] (o4) {$D_{i}, F0_{i}$};
\path [line] (in4) -- (E1);
\path [line] (E1) -- (o4);

\node[block,node distance=1.1cm, minimum height=2em, below of=E1] (D) {$\mathbb{E}(StCSH)$ Estimation};
\node[io, left of= D] (in5) {$\{SC_{i}\}_{i=1}^{N}$};
\node[io,right of=D] (o5) {\textit{d}};
\path [line] (in5) -- (D);
\path [line] (D) -- (o5);

\node[block, node distance=1.1cm, minimum height=2em, below of=D] (E) {Feature Extraction II};
\node[io, left of= E] (in6) {$SC_{i},$ \textit{d}};
\node[io,right of=E] (o6) {$CSHsc_{i}$};
\path [line] (in6) -- (E);
\path [line] (E) -- (o6);
\end{tikzpicture}
\caption{Block scheme of the proposed method. The $1^{\mathrm{st}}$, $3^{\mathrm{rd}}$, $4^{\mathrm{th}}$ and $5^{\mathrm{th}}$ block are intended to be embedded in a for loop over the elements of the dataset (i.e. $i=1, \ldots N$). $SC_{i}^{0}$ is the first segmentation of the signal given the $i^{\mathrm{th}}$ couple cry-noise ($C_{i}, N_{i}$); $\overline{\textit{r}}$ is the lower confidence limit for the mean of the spectral entropy of $N_{i}$; $\hat{q}_{.85}$ is the upper basic bootstrap confidence limit for the $0.85$-quantile of the permanence time of spectral entropy of $C_{i}$ under the threshold $\overline{\textit{r}}$; $SC_{i}$ is the final segmentation of the  $i^{\mathrm{th}}$ cry signal; \textit{d} is the lower 0.95-confidence limit for the mean of the random variable \textit{StCSH}, defined in (1); $ D_{i}, F0_{i}$ and $CSHsc_{i}$ are the proposed LOD indicators.
In Section 4 we do not use the \textit{i} indexes for the sake of simplicity.}
\label{Flow}
\end{figure}
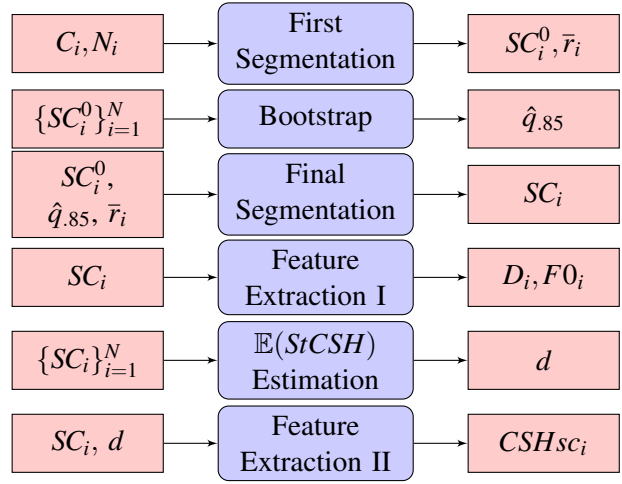

\subsection{Pre-processing}
After being extracted from the video, every audio track has been cut manually: for every infant, we considered the $30s$ after the painful stimulus \cite{facchini2005relating}. In what follows we will refer to each of these $30s$-long signals as cry signal ($C_{i}$ in Fig. \ref{Flow}).
First, we checked the dataset for audio tracks containing non-stationary background noises (e.g. speech signals, other crying infants etc$\ldots$) that may interfere with the vocal expression and removed them. Then, in order to initialise the method, we inspected the original audio records for an interval of at least $1s$ of stationary background noise ($N_{i}$ in Fig. \ref{Flow}) located as close as possible to the occurrence of the painful stimulus. Because of bad recording environment, this has not been always possible and all those cry signals without an associated background noise have been discarded. The final dataset consists of 14 cry signals coupled with 14 
background noise samples: $(C_{i}, N_{i})$, with $i=1, \ldots 14$.

These coupled data have been used as input to an R \cite{R} script. The required audio analysis tools are part of the package 'seewave' \cite{Seewave} and 'tuneR' \cite{tuneR}. The procedure consists of six blocks (Fig. \ref{Flow}): the first three of them concur in the segmentation procedure and the last three blocks perform the feature extraction.

\subsection{Segmentation}

\begin{figure}[t]
\centering
{
\includegraphics[width=3.1in]{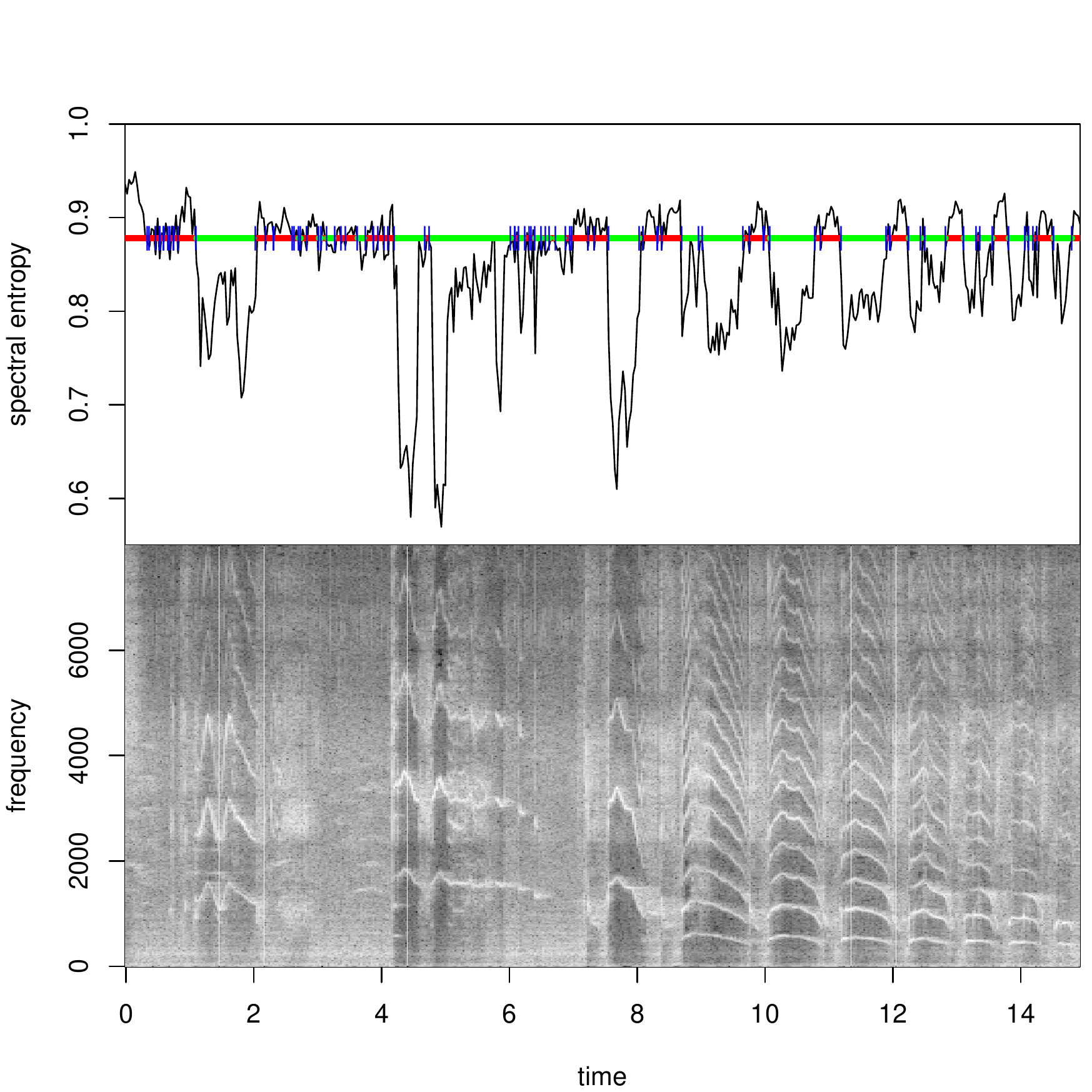}
\caption{A comparison between \textit{HC} and the spectrogram of the cry signal. The horizontal line is the corresponding noise threshold $\overline{r}$. The green intervals are classified as cry units in the first segmentation and the vertical bars {\color{blue}| }represent the respective cut-points.\label{cshvsspec}}
}
\end{figure}
An automatic segmentation algorithm is a primary step towards infant cry analysis. In fact, in order to evaluate the features of a given cry signal, we need to first be able to detect every continuous interval of time in which a vocalization (i.e. a product of the infant's vocal apparatus different from the ihalation-related sounds) occurs. We will refer to these intervals as cry units, although the precise meaning of this term will be defined numerically at the end of this section.

We have treated this detection problem by considering the continuous spectral entropy (CSH) of the signal \cite{toh2005spectral}(Ap\-pen\-dix). Given a couple of signals in the dataset, let us call \textit{HN} the CSH of the stationary noise sample and \textit{HC} the CSH of the cry. By considering the lower confidence limit for the mean of \textit{HN} (let us call it $\overline{\textit{r}}$), we have observed that (Fig. \ref{cshvsspec}):
\begin{itemize}
\item The entropy of the inhalation related sounds is above $\overline{r}$;
\item When \textit{HC} is lower than $\overline{r}$, it describes some U-shaped patterns of variation, and the most relevant of them corresponds to a voiced pattern in the spectrogram.
\end{itemize}

\begin{figure}[t]
\centering
{
\includegraphics[width=3.1in]{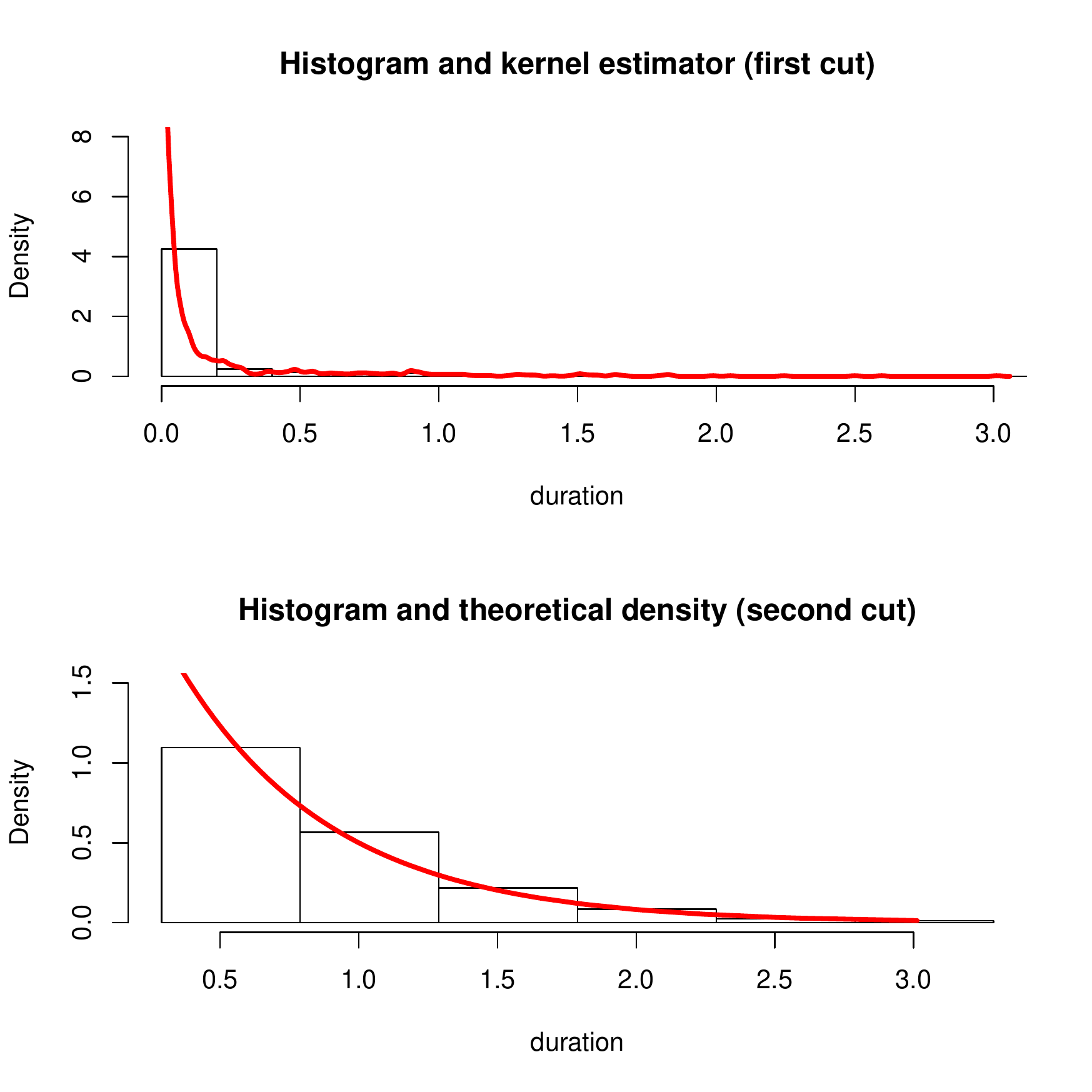}
\caption{The histograms of the intervals' durations after the first (top) and the second (bottom) segmentation of the cry. The second cut  considers only those intervals longer than $\hat{q}_{.85}$, that is the upper tail of the firs cut.\label{CutTau}}
}
\end{figure}

Thus, by considering all those time intervals such $\textrm{\textit{HC}}<\overline{r}$, we obtained a first segmentation of the cry signal ($SC_{i}^{0}$ in Fig. \ref{Flow}). Now, if performed over the whole dataset, this procedure returns 1352 time intervals with a mean duration of about $0.13s$. This value is far too small from a psychoacoustics point of view: indeed, the human perception of sounds starts to change dramatically under a duration threshold equal to $512ms$ \cite{gelfand2016hearing}. To increase this value, we need to remove from this segmentation all those time intervals whose duration is not significantly long. From a stochastic point of view, this means to find some kind of boundary for the permanence time of \textit{HC} under the threshold $\overline{r}$. Let us call $\tau$ this permanence time, which, in this method, represents also the random variable "duration of a cry unit". The first segmentation returns 1352 realizations (call them $\hat{\tau}$) of $\tau$. Looking at the empirical density function of $\hat{\tau}$, we noticed that the duration of more than the $90\%$ of the intervals is less than $0.5s$ (Fig. \ref{CutTau}). Moreover, we observed that many of these brief time intervals correspond to the occurrence of some non-stationary or transient noise in the original signal, while the rare long-lasting ones correspond to exceptionally long cry units. In particular, for those records in which there is just a brief moan or no cry at all, this first cut fragments the signal in a set of very short segments, returning more noisy intervals than cry units. So, we obtained a second and final segmentation of the signal by removing from the first one all those intervals whose duration is less than the upper basic bootstrap confidence limit \cite{canty2012boot} for the $0.85$-quantile\footnote{The more common 0.9-quantile results in a cutoff that exceeds the $0.5s$ empirical threshold of approximately $43ms$.} of $\tau$, which determines a cutoff $\hat{q}_{.85}$ of about $288ms$.

In the resulting segmentation, those signals containing just brief moans become almost silent and all the most significant cry units are preserved in all the cry episodes. Moreover, among all the 166 intervals (with average duration of about $844ms$) identified in this way, just one contained pure noise and the others were exact cry units. The duration of these time intervals can be modeled (K-S statistic $\textrm{D}=7.8\cdot 10^{-2}$, \textit{p}-value$=0.2$) as $X + \hat{q}_{.85}$, where $X$ is an exponential with maximum likelihood estimated \cite{fitdistrplus} rate parameter 1.8 (Fig. \ref{CutTau}).

\subsection{Feature Extraction}
In this second part of the process, we use the segmented cry (\textit{SC}) to measure some distress-related features of the original cry. As we said, some cry signals can result in a \textit{SC} which is totally silent. We set to $0$ all the feature-related scores for these signals. In what follows we denote with $M$ the number of cry units in the \textit{SC} and suppose $M\geq 1$.

\begin{figure}[t]
\centering
\includegraphics[width=3.1in]{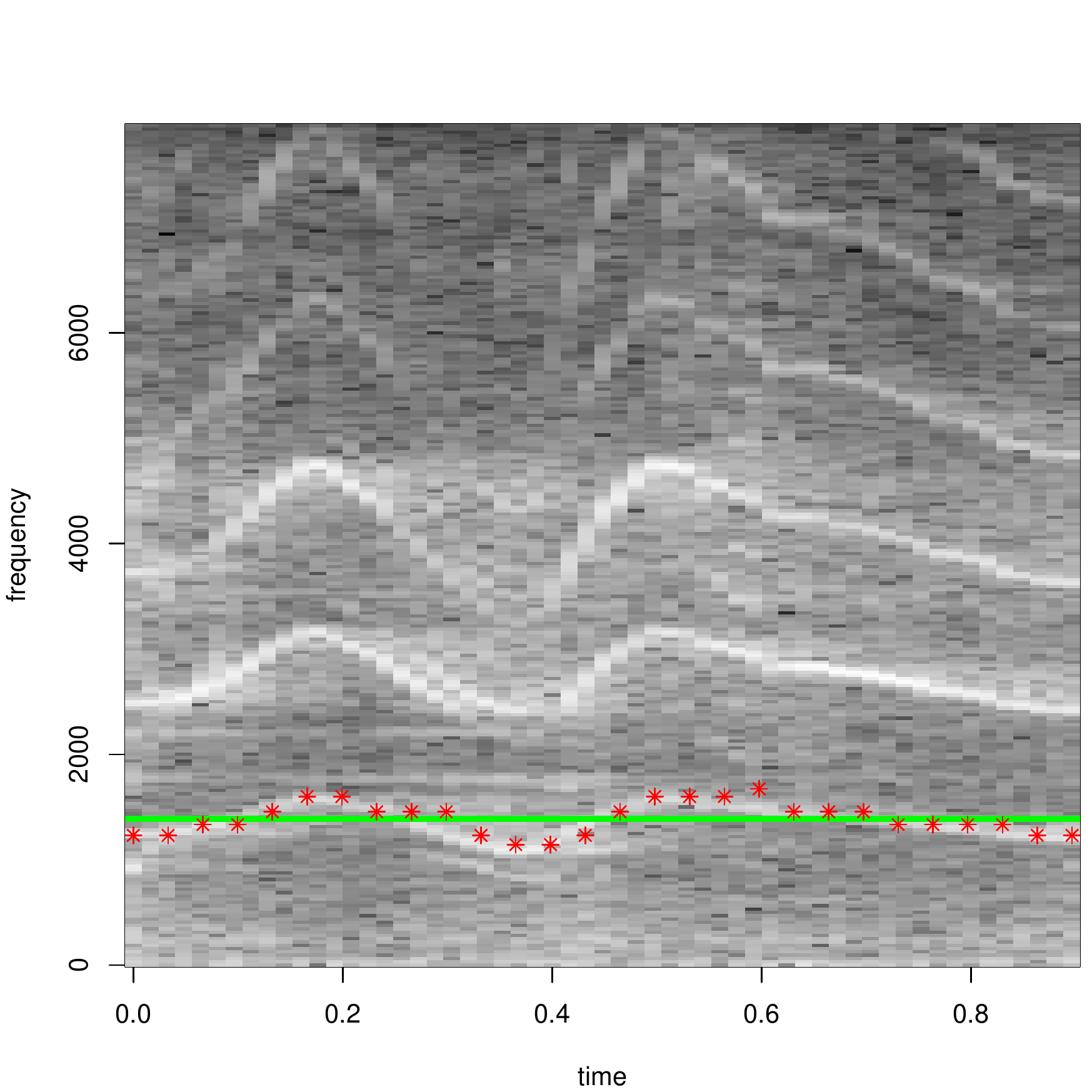}
\caption{Spectrogram of the first significant cry unit, \textit{FC}. Each {\color{red}$\ast$} represents the fundamental frequency in the respective window estimated by the corresponding high-frequency ripple in the cepstrogram of \textit{FC}. The green line is the mean fundamental frequency of the first cry \textit{F0}.\label{FC}}
\end{figure}

\subsubsection{Duration}
The total duration of the most significant cry units in the 30 seconds after the painful stimulus (\textit{D}) is undoubtedly an interesting characteristic of the cry signal. In fact, observe that the vocal expression item of the DAN scale \cite{carbajal1997dan} is actually an evaluation of the duration of the cry. If we denote with $\textbf{s}$ and $ \textbf{e}$ the $M$-dimensional vectors recording the cry units' starting and ending points respectively, then:
\[\textrm{\textit{D}}=\sum_{i=1}^{M}(e_{i}-s_{i}).\]
So, \textit{D} can be calculated just after the final segmentation (fourth block in Fig. \ref{Flow}).
\subsubsection{Fundamental frequency of the first cry}
As already said in the Section 1, one other outstanding feature of the cry is its fundamental frequency, and in particular the fundamental frequency of the first cry after a painful stimulus \cite{baeck2001study,porter1986neonatal}. The \textit{SC} provides us not the exact first cry but the first significant cry unit after the heel lance, which is the signal in the time interval $[s_{1}, e_{1}]$, if \textit{SC} is not completely silent (i.e. $M\geq 1$). Let us call first cry (\textit{FC}) this signal. Now, by considering the high-frequency ripples \cite{noll1967cepstrum} in the cepstrogram (Appendix) of \textit{FC} we can provide a series (with sample frequency equal to the ratio between the window length, 512 in our model, and the sample frequency) of estimates of the fundamental frequency. Thus, we will approximate the fundamental frequency with the sample mean of this series (Fig. \ref{FC}): we will call \textit{F0} this quantity, which is also calculated after the final segmentation (fourth block in Fig. \ref{Flow}).

\subsubsection{CSH score}
The energy of a signal is as unstructured as it is dispersed on a larger range of frequencies \cite{shannon2001mathematical} i.e. as its spectral entropy is near to 1. Clearly the CSH of the \textit{SC} will never be equal to 1, because it is bounded by the threshold $\overline{\textit{r}}$ (that depends on the noise with which each cry is coupled) that we used to obtain the first cut. Thus, in order to construct a scoring system equal for every cry signal, we classified a certain time window in a cry unit as "dysphonated" if, in that time window, its CSH is significantly close to the noise threshold $\overline{\textit{r}}$. The specification of how much the distance between these two quantities has to be near to $0$ is a thresholding problem involving the random variable (r.v.) \textit{StCSH}, defined as:
\begin{equation}
\textrm{\textit{StCSH}} = \overline{\textit{r}} - H(\textrm{windowed cry unit}), 
\end{equation}
where $H$ denotes the spectral entropy (Appendix). Now, let us suppose that we have $N$ cry signals: $C_{1}, \ldots C_{N}$ such that all of them give us a $\textrm{\textit{SC}}_{i}$ with $i=1, \ldots N$ which is non totally silent. For $i=1, \ldots N $, let $\textbf{s}_{i}$ and $\textbf{e}_{i}$ be the $M_{i}$-dimensional vector containing the starting and ending points of the cry units in $\textrm{\textit{SC}}_{i}$ respectively. We denote with $\overline{r}_{1}, \ldots \overline{r}_{N}$ the respective noise thresholds in the CSH. Let us consider a Hanning's window 
\[w(t)=\Big[\frac{1}{2}+\frac{1}{2}\cos \Big(\frac{2\pi}{a} t\Big)\Big]\one_{[-\frac{a}{2}, \frac{a}{2}]}(t),\]
where $a>0$ is arbitrary but fixed and $\one_{[-\frac{a}{2}, \frac{a}{2}]}$ is the indicator function. By using the traslation operator $\vartheta_{s}: w(t)\mapsto \vartheta_{s}w(t) = w(t-s)$ to slide the window, then we obtain a realization 
\[\textrm{\textit{StCSH}}_{ij}(s) = \overline{r}_{i} - H(\vartheta_{s}w \hspace{1mm}C_{i}\hspace{1mm}\one_{[s_{ij}, e_{ij}] })\]
of the r.v. \textit{StCSH} for every $s$ in the support of $\one_{[s_{ij}, e_{ij}]}$, for every $j=1, \ldots M_{i}$ and for every $i=1, \ldots N$. The discrete nature of the signal (and the Heisenberg-Pauli-Weyl uncertainty inequality \cite{grochenig2013foundations}) forces us to consider only a finite set of equispaced istants of time. So, given a cry unit $C_{i}\hspace{1mm}\one_{[s_{ij}, e_{ij}]}$, we considered only the the realizations of \textit{St\-CSH} corresponding to the points $t_{1}, \ldots t_{K_{ij}}$:
\begin{equation*}
	s_{ij}<t_{1}=s_{ij}+\frac{a}{2}<t_{2}=t_{1}+a<\ldots<t_{K_{ij}-1}<t_{K_{ij}}=s_{ij}+a K_{ij}\leq b
\end{equation*}

where $K_{ij}$ is the integer part of $(e_{ij}-s_{ij})/a$.
We used those realizations to give an estimate \textit{d} of the lower 0.95-confidence limit for the mean of \textit{StCSH}.

Fixed the threshold \textit{d}, we can assign a dysphonation score to every cry by counting all the time windows in which the distance between the CSH of the \textit{SC} and the corresponding $\overline{\textit{r}}$ is less than \textit{d} (the last two blocks in Fig. \ref{Flow}). We named CSH score (\textit{CSHsc}) this quantity multiplied for the length of the window. More formally, we gave the following:

\paragraph{Definition}{Given a cry signal $C_{i}$ and a segmentation $\textrm{\textit{SC}}_{i}=\{[s_{ij},e_{ij}]\}_{j=1, \dots M_{i}}$, chosen a Hanning's window of length $a$, we define:}
\begin{equation*}
\textrm{\textit{CSHsc}}_{i}=\left\{
\begin{array}{lll}
a\sum_{j=1}^{M_{i}}\sum_{h=1}^{K_{ij}}\one_{\{ \textrm{\textit{StCSH}}_{ij}<\textrm{\textit{d}}\}}(t_{h}) & \mbox{if} & M_{i}\geq1,\\
0 & \mbox{if} & M_{i}=0.
\end{array}
\right.
\end{equation*}

As we said, the final segmentation gives us 166 cry units for a total duration of approximately $137s$. Then, a Hanning's window with $a=32ms$, by sliding along these cry units, produces about 4281 realizations (that we assume to be independent) of \textit{StCSH}, with estimate $\textit{d}\approx 77\cdot 10^{-3}$ of the lower 0.95-confidence limit for its mean.

\begin{figure}[t]
\centering
\includegraphics[width=3.1in]{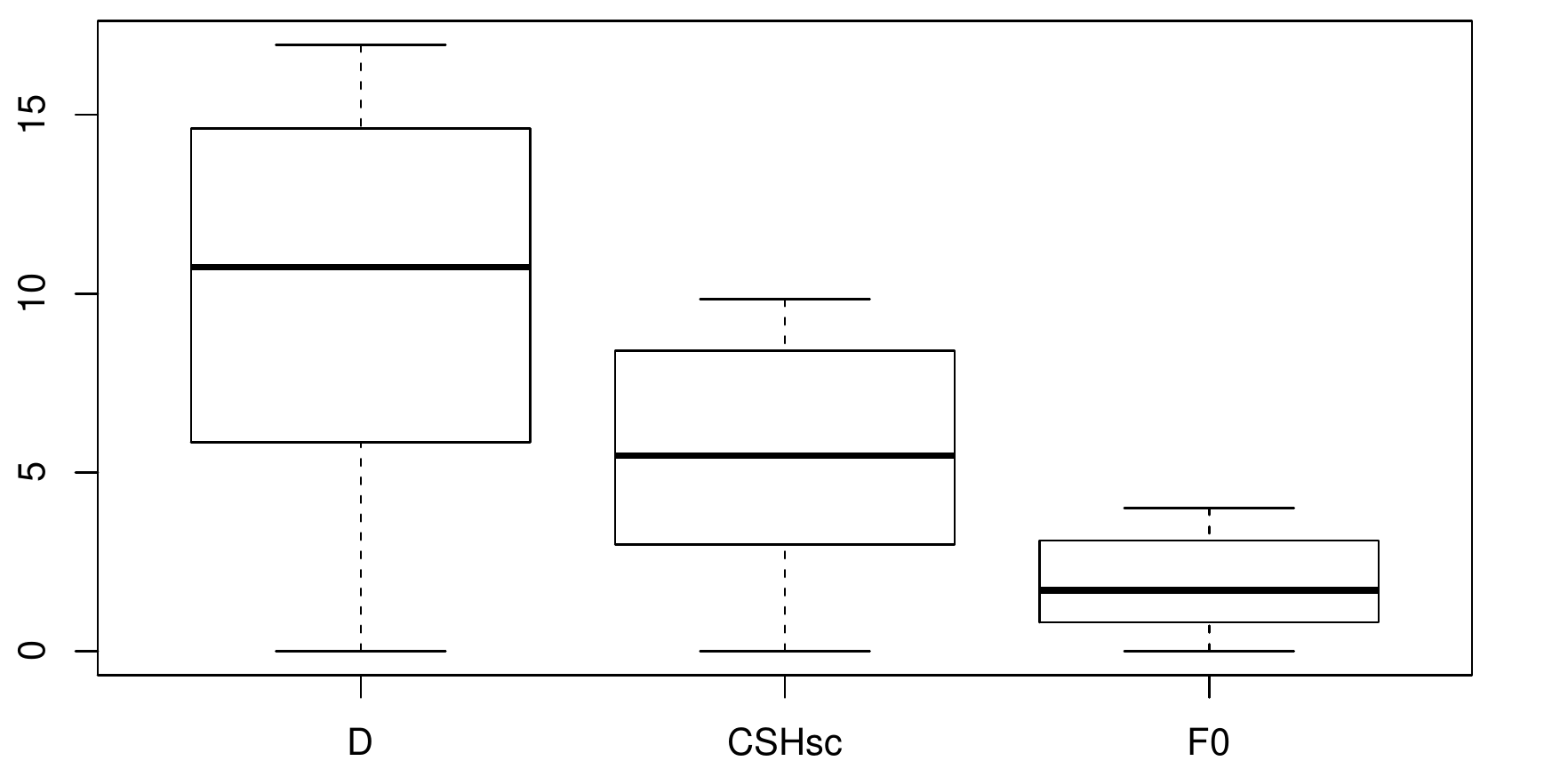}
\caption{Box-plots of the proposed indicators \textit{D}, \textit{CSHsc} and \textit{F0}.}
\label{box}
\end{figure} 

\section{Results}
In this section we analyze the output of the overall algorithm for the given dataset and in particular the three extracted features/scores: \textit{D}, \textit{CSHsc} and \textit{F0} (box-plots in Fig. \ref{box}). 

\subsection{Variables' correlation} 
The couple \textit{D-CSHsc} is correlated (estimated Pearson's correlation coefficient $\varrho=0.84$, \textit{p}-value$=1.4 \cdot 10^{-4}$). This result is in agreement with the fact that both the duration and \textit{CSHsc} have been constructed on the CSH-derived segmentation of the cry, resulting in an inner common dependency by the CSH of the signal and its noise threshold $\overline{\textit{r}}$. A more interesting fact is that \textit{F0} is correlated with both \textit{D} ($\varrho=0.74$, \textit{p}-value$=2.3\cdot 10^{-3}$) and \textit{CSHsc} ($\varrho=0.82$, \textit{p}-value$=3.3\cdot 10^{-4}$), and that this latter correlation is actually greater than the former one. 

\begin{figure}[!t]
\centering
\includegraphics[width=3.1in]{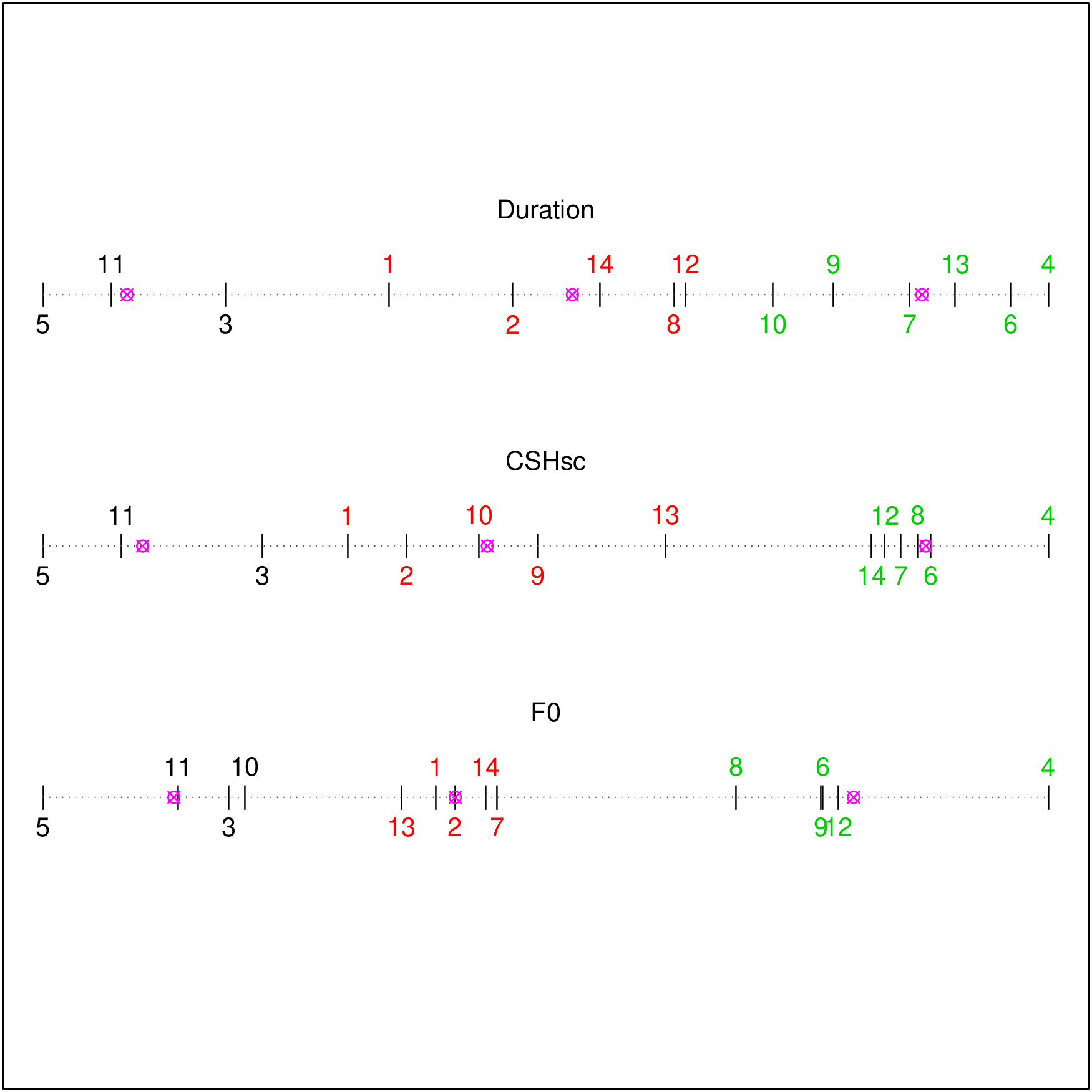}
\caption{A representation of the 3-means clusters for the proposed indicators. The elements of the dataset are labeled with numbers whose color identifies the cluster. The clusters'centers are denoted by \raisebox{+0.4ex}{{\color{magenta}{\tiny{$\otimes$}}}}.}
\label{cluster}
\end{figure}

Now, because the intended use of these indicators is to construct a machine-based pain evaluation tool, a preliminary step is to check if their values highlight the presence of the three levels considered by the most common pain assessment scales (i.e. "Mild", "Moderate", "Severe"). Thus, we have turned the continuous scores in categorical data via unidimensional 3-means clustering \cite{wang2011ckmeans} (Fig. \ref{cluster}). We have compared the resulting classifications by considering their contingency tables and performing the Pearson's chi-squared test on each of them. The results are reported in Table 2. The Spearman's $\rho$ rank correlation helps us to understand if the correlation between the continuous indicators is still present in the paired classifications. The null hypothesis of independence is rejected for all the couples, even though  only \textit{CSH\-sc} and \textit{F0} are strongly correlated.

\begin{table}[h]
\renewcommand{\arraystretch}{1.3}
\centering
\caption{Pearson's $\chi^{2}$ test outcomes and Spearman's $\rho$ rank correlation estimaes}
\begin{tabular}{l | c c || c c}
Couple & $\chi^{2}$ & \textit{p}-value & $\rho$ & \textit{p}-value \\
\hline
\textit{D-CSHsc} & 14.14 & $0.6\cdot 10^{-2}$ & 0.54 & $4.5\cdot 10^{-2}$ \\
\textit{CSHsc-F0} & 12.13 & $1.6\cdot 10^{-2}$ &$0.77$& $0.1\cdot 10^{-2}$\\
\textit{D-F0}& 10.43 & $3.4\cdot 10^{-2}$& 0.5& $6.5\cdot 10^{-2}$\\
\end{tabular}
\end{table}

\subsection{Human-based assessment} 
Because the expected receiver of the infant's cry is a human listener, we considered 6 human-based assessments of the same dataset. The scorers were two near-graduate students ($S_{a}$ and $S_{b}$) in pediatric nursing, who repeated the assessment twice ($t_{0}$ and two months later, $t_{1}$) and two experienced pediatric nurses ($GS_{a}$ and $GS_{b}$). Each of them was provided with the audio-video record of the heel lance and was asked to assess the pain by using the DAN scale (Table 1). We only considered the score of the "Vocal expression" item. Moreover, in order to make the comparison feasible with the 3-mean cluster classifications (Fig. 6), we identified the first two scores of the "Vocal expression" item (i.e. both "No complaints" and "Moans briefly" are labelled with 1). At the beginning we grouped the scorers as follows:
\begin{itemize}
\item Group I: $S_{a}^{t_{0}}$, $S_{a}^{t_{1}}$;
\item Group II: $S_{b}^{t_{0}}$, $S_{b}^{t_{1}}$;
\item Group III: $GS_{a}$, $GS_{b}$.
\end{itemize}
To quantify the correlation of the evaluations, we considered the respective contingency tables for each group and performed the Pearson's chi-squared test. The null hypothesis of independence is rejected for all the couples (\textit{p}-value$<0.04$). The reiterated evaluations are the most correlated ($\rho\approx0.9,0.89$ for $S_{a}$ and $S_{b}$ respectively), while $\rho\approx0.66$ for the couple $GS_{a}$- $GS_{b}$. 

We carried out a between-groups analysis in order to understand if it is suitable to use an experienced observer as gold standard for pain assessment. First we have turned both  $GS_{a}$ and $GS_{b}$ in binary classificators by partitioning the possible outcomes in equal or strictly lower than 3. Then, one at a time, the resulting Boolean variables have been used as correct classification to construct one confusion matrix for every other observer in the dataset. We evaluated the performance of each classification with the ROC curve and in particular the area under the curve (AUC) \cite{ROCR2005}. The results of this analysis indicate that the assessment of the experienced observers ($0.67\leq \textrm{AUC}\leq 0.98$) is not meaningfully different from the scores of the inexperienced ones ($0.65 \leq \textrm{AUC} \leq 0.95$).

We performed the same kind of analysis on the final DAN score (Table 1). This analysis required to fix a cutoff for the DAN scale in order to turn the assessments of the experienced observers in Boolean variables. We tried different values: the resulting AUC did not show any kind of improvement or worsening pattern in dependence by the choice of this cutoff.

\subsection{Correlation between human-based assessment and output variables}
The values of the AUC do not identify a difference in the performance of the three groups of human observers. So, to choose a scorer and use its assesment as correct classification seems quite arbitrary in this scenario. Therefore, we considered all the 6 human-based assessments as realizations of the variable "Human Scorer" aiming to compare the human-based assessment of the infant's vocal expression to the values of the output variables. After excluding all those cases (2 in the dataset, Table 3) which are not unimodal, we considered the sample mode of these 6 human-based assessments of the vocal expression (\textit{MoH}).

\begin{table}[t]
\centering
\caption{Human observers-Indicators}
\scalebox{0.8}{
\begin{tabular}{c | c c c | c c c}
& & Score Frequecy & & & Indicators & \\
\Tstrut Label & I & II & III & D & CSHsc & F0  \Bstrut \\
\hline
1 & 0 & 4 & 2 & 5.83 & 2.98 & 1.56 \Tstrut \\
2 & 3 & 2 & 1 & 7.92 & 3.56 & 1.64 \\
3& 4 & 2 & 0 & 3.08 & 2.15 & 0.74 \\
4& 0 & 0 & 6 & 16.96 & 9.84 & 4 \\
5& 6 & 0 & 0 & 0 & 0 & 0 \\
6& 0 & 0 & 6 & 16.31 & 8.69 & 3.1 \\
7& 0 & 2 & 4 & 14.62 & 8.4 & 1.81 \\
8& 0 & 3 & 3 & 10.64 & 8.56 & 2.76 \\
9& 0 & 3 & 3 & 13.33 & 4.84 & 3.1 \\
10& 0 & 4 & 2 & 12.31 & 4.26 & 0.8 \\
11& 4 & 2 & 0 & 1.15 & 0.77 & 0.54 \\
12& 0 & 0 & 6 & 10.83 & 8.24 & 3.16 \\
13& 0 & 5 & 1 & 15.38 & 6.09 & 1.43 \\
14& 0 & 0 & 6 & 9.39 & 8.11 & 1.76 \\
\end{tabular}
}
\end{table}

Thus, we constructed the contingency tables between the 3-mean clusters of the proposed indicators and \textit{MoH} for the unimodal cases. Again we performed the Pearson's chi-squared test on each of them and calculated the Spearman's $\rho$ rank correlation of every couple of classifications. The best result is given by the \textit{CSHsc} classification for both Pearson's chi-squared test ($\chi^{2}=18.75$, \textit{p}-value $=8.8\cdot 10^{-4}$) and Spearman's rank correlation ($\rho= 0.96$), while for the others the null hypothesis of Pearson's chi-squared test is not rejected (\textit{p}-values $=7\cdot 10^{-2}$ for both \textit{D} and \textit{F0}). By interpreting the class of \textit{MoH} as levels, we can try to fit them with a linear model of the form 
\begin{equation}
MoH = \alpha_{1} \textrm{D} + \alpha_{2} \textrm{CSHsc} + \alpha_{3} \textrm{F0}+ \epsilon.
\end{equation}
Estimating the coefficients in (2) for the given dataset, \textit{CSHsc} results to be not only the most relevant indicator ($\alpha_{2} =0.34$, std. error $=0.9$ \textit{t}-value $=3.7$, \textit{p}-value $=5.5\cdot 10^{-3}$), but also the only one with a coefficient significantly different from $0$ (\textit{p}-value $>0.36$ for both \textit{D} and \textit{F0}).

\begin{figure}[t]
\centering
\includegraphics[width=3.1in]{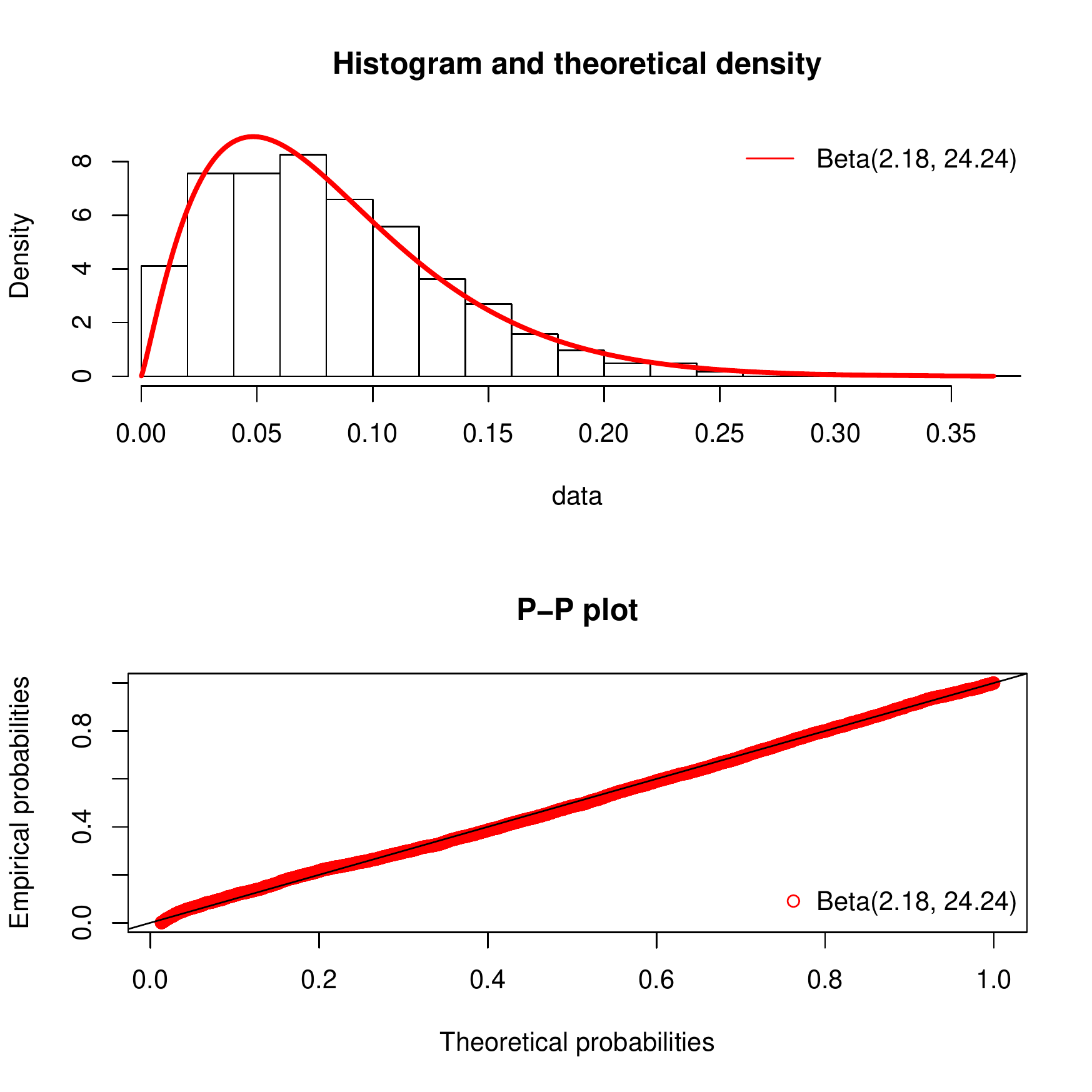}
\caption{A comparison between the theoretical and empirical densities and cumulative density functions of the $5\%$-trimmed variable \textit{StCSH}.\label{denscomp}}
\end{figure} 

\subsection{The StCSH variable}
Let us consider \textit{StCSH} defined by (1). Because \textit{CSH} is standardized by subtracting the corresponding noise threshold $\overline{r}$, it is reasonable to think that the lowest values of \textit{StCSH} contain outliers. Thus, we considered the lower $5\%$-trimmed empirical distribution of \textit{StCSH} for the given dataset. In particular, we observed that a beta r.v. with maximum likelihood estimated parameters 2.18 and 24.24 fits the observed values of \textit{StCSH} (K-S statistic $\textrm{D}=1.9\cdot 10^{-2}$, \textit{p}-value$=9.5\cdot 10^{-2}$, Fig. \ref{denscomp}).

\section{Discussion and conclusions} 
%This paper is part of a preliminary investigation commissioned by AO Ordine Mauriziano di Torino (Italy) to Istituto Superiore Mario Boella. Given a small amount of data, the aim of the project was to explore machine-based methods for infant pain-assessment in order to study the  feasibility of an automatic pain assessment tool. 
In the context of a feasibility study on the development of an automatic pain assessment tool, given a small dataset of cry-noise coupled signals, we have provided a basic method to compute three correlated LOD indicators. 

All the estimates were calculated with non-parametric setting. Moreover, differently from other infant cry analysis procedures, the proposed method is entirely implemented on the R \cite{R} free software, making all its steps completely specifiable by an accessible source code and therefore reproducible.

The preliminary nature of the study forced us to operate without a big dataset of infant cry, which is one of the novelty of the proposed method. In fact, as far as we know, the majority of the methods in the literature relies on the use of models whose parameters have to be trained (e.g. Hidden Markov \cite{xie1996automatic}, \cite{abou2015use}, Random Forest \cite{orlandi2015application}), therefore requiring a big amount of data. Besides the peculiar context of this study, to assemble a database of infant cry is not an easy task as there are multiple difficult aspects to take into account to develop of such database (see \cite{chittora2017data} for details about the ideal characteristics of an infant cry corpus):
\begin{itemize}
\item Technical: install and use an audio acquisition framework in a neonatal unit, which is a noisy and uncontrolled environment;
\item Legal: confidentiality, parental consensus and privacy;
\item Standardization: once acquired, each cry has to be labeled by the context (e.g. distressfull, painful etc.). Moreover, the acoustic features have a great variability with age, weight and gestational age etc.
\end{itemize}
Each of these aspects becomes even more difficult in the case of the most immature and sick infants, whose pain has to be reliably assessed in order to be menaged. From this point of view, a method based on the use of a big dataset of cry signals could be very impractical, unless it leads to outstanding performances.
 
On the contrary, the proposed method relies only on the statistical properties of the spectral entropy of the cry. Observe that the use of windows, sliding across a small amount of signals, provided us with statistically significant samples to estimate these properties. So, we have built a method despite the small amout of data, which is applicable even with a dataset of just one cry coupled with a stationary noise. However, once we are able to collect a bigger dataset of both recordings and evaluators, our next objective will be the comparative study of the proposed method with the existing ones in terms of output and performances. 

Among the proposed LOD indicators, \textit{D} and \textit{F0} are well-known: the duration and the fundamental frequency of the first cry are considered meaningful in the literature \cite{baeck2001study} and their computation is easy, once the cry segmentation is given. It is worth to say that these two features have been selected among a greater set of possible distress-related characteristics of the cry. Other indicators suggested in the literature (e.g. the variance in the cry units duration \cite{baeck2003bayesian} or in the fundamental frequency during the cry \cite{porter1986neonatal}, the root mean square of the time wave \cite{bellieni2004cry}) displayed poor correlation between the other extracted features (estimated Pearson's correlation coefficient $\varrho<0.2$) and therefore their evaluation has been removed from the method. 

The CSH score, as far as we know, provides a new way to measure the dysphonation of the cry by tracking the presence of unstructured energy in it. The "dysphonation phoneme" was introduced by  Xie et al. \cite{xie1996automatic} as state in a Hidden Markov Model (HMM). In particular, it was observed \cite{xie1993determining} that the permanence time in the dysphonation state shows the most consistent positive correlation with the perceived LOD. The dysphonation phoneme is characterized by "an unstructured energy distribution over all the frequency range, sometimes with a tendency of higher concentration over the middle to high (1-5 kHz) frequency range or an unstructured energy distribution imposing on or in between the barely distinguishable harmonics"\cite{xie1993determining}. This is the definition of just one out of the 10 states of the HMM proposed in \cite{xie1996automatic}, each of them is a phoneme analogously described by a time-fre\-quen\-cy pattern of variation. The training of a such HMM requires a lot of data and effort. So, instead of applying this  HMM to find an estimate of the permanence time in just one of the 10 states of the model, we preferred to track and measure the occurrence of the dysphonation phoneme in the segmented cry by monitoring the presence of unstructured energy in it. Besides this operative practicality, the \textit{CSHsc} is higly correlated with both \textit{D} and \textit{F0}. Moreover, it can be modeled as the permanence time of a process with known distribution under a threshold, making this new indicator particularly interesting for further investigations. The most relevant result is that the 3-mean clusters of \textit{CSHsc} are highly correlated with the sample mode of the human scorers making it a candidate predictor in a hypothetical model for the human-based assessment of LOD of the infant's vocal expression. 

Because of the significant correlation of the proposed indicators when considered as continuous variables, our purpose would be using them as predictors of the human-based assessment of the LOD via a general linear model (an ordered logit would be suitable, in our opinion). Clearly this validation process requires a bigger dataset of both recordings and evaluations, therefore more data are needed.  

\appendix
\section*{Appendix}
\subsection*{Continuous Spectral Entropy} Let $X$ be a discrete random variable (d.r.v.) such that
\begin{align*}
&\mathbb{P}(X=n)=\mathbb{P}(A_{n})=p_{n}, &n\in \mathbb{N}.
\end{align*}
The \textit{spectral entropy} of $X$ \cite{shannon2001mathematical} is defined as:
\[H(X)=-\sum_{n\in \mathbb{N}}p_{n}\log p_{n}.\]

Given a discrete signal $\textbf{y}\in \mathbb{C}^{N}$, let us denote with \[\{F_{n}\textbf{y}\}_{n=0, \ldots N-1}\] its discrete Fourer transform. Then, thanks to the discrete Plan\-che\-rel's equality \cite{gasquet2013fourier}, we can define the d.r.v. $S_{\textbf{y}}$ such that:
\begin{align*}
&\mathbb{P}(S_{\textbf{y}}=n)=\left\{\begin{array}{l l}
\frac{|F_{n}\textbf{y}|^{2}}{N\Vert\textbf{y}\Vert^{2}} & \mbox{ if $n=0,\ldots N-1$;}\\
0 & \mbox{ if $n\geq N$.}
\end{array} \right.
\end{align*}

The \textit{spectral entropy} of \textbf{y}$\in\mathbb{C}^{N}$ is defined as:
\[H(S_{\textbf{y}})=-\sum_{n=0}^{N-1}\frac{|F_{n}\textbf{y}|^{2}}{N\Vert\textbf{y}\Vert^{2}}\log\frac{|F_{n}\textbf{y}|^{2}}{N\Vert\textbf{y}\Vert^{2}}.\]

By calculating $H$ for every element in the spectrogram of \textbf{y}, i.e. for $S_{\textbf{yw}_{0}}, \ldots S_{\textbf{yw}_{M}}$ where $\textbf{w}_{j}$ is a discrete sliding window, we get the \textit{continuos spectral entropy} of \textbf{y} \cite{toh2005spectral}.

\subsection*{Cepstrogram} 
Given a discrete signal $\textbf{y}\in \mathbb{C}^{N}$, let us define: \[\textbf{z}=(\log|F_{0}\textbf{y}|, \ldots \log|F_{N-1}\textbf{y}|).\]
Then the \textit{discrete cepstrum} \cite{noll1967cepstrum} $C\textbf{y}\in\mathbb{R}^{N}$ of \textbf{y} is defined as
\begin{align*}
& C_{k}\textbf{y}=\Re (F^{-1}_{k}\textbf{z}) & k=0,\ldots N-1.
\end{align*}

In analogy with the spectrogram, the \textit{cepstrogram} of \textbf{y} is defined as the cepstrum of the windowed signal: $\{C\textbf{yw}_{j}\}_{j=1, \ldots M}$ where $\textbf{w}_{j}$ is a discrete window.

\bibliographystyle{IEEEtrans}
\bibliography{bibfile}   
\end{document}